\documentstyle[12pt]{article}

\begin{document}
\begin{titlepage}
\begin{center}
\today     \hfill    LBL-31711 \\
           \hfill    UCB-PTH-92/05 \\

\vskip .5in

{\large \bf The 2-Dimensional Quantum Euclidean Algebra}
\footnote{This work was supported in part by the Director, Office of
Energy Research, Office of High Energy and Nuclear Physics, Division of
High Energy Physics of the U.S. Department of Energy under Contract
DE-AC03-76SF00098 and in part by the National Science Foundation under
grant PHY90-21139.}

\vskip .5in

Peter Schupp, Paul Watts, and Bruno Zumino \\[.5in]

{\em Department of Physics\\
     University of California\\
     and\\
     Theoretical Physics Group\\
     Physics Division\\
     Lawrence Berkeley Laboratory\\
     1 Cyclotron Road\\
     Berkeley, California 94720}
\end{center}

\vskip .5in

\begin{abstract}

The algebra dual to Woronowicz's deformation of the 2-\-di\-men\-sion\-al
Euclidean group is constructed. The same algebra is obtained from $SU_{q}(2)$
via contraction on both the group and algebra levels.

\end{abstract}
\end{titlepage}

\section{Introduction}

The Euclidean group $E(2)$ is a simple example of an inhomogeneous group.
Deformations of such groups in general have been studied in
\cite{SWW}.  Celeghini et al. \cite{CGST} found a deformation of $Ue(2)$ by
contracting $U_{q}su(2)$ and simultaneously letting the deformation
parameter $h \equiv \ln q$\, go to zero. Here we are interested in the
case where $q$ is left untouched.
Elements of the general theory of quantum groups can be found in \cite{RTF}
and references therein.

\section{$U_{q}e(2)$ as the dual of $Fun(E_{q}(2))$}
\label{sec-Paul}

In this section, the dual to Woronowicz's deformation of $E(2)$ \cite{W} is
constructed explicitly, using techniques similar to those of Rosso \cite{R}.
Woronowicz introduces Hopf algebra elements $n$, $v$,
$\overline{n}$, and
$\overline{v}$, which satisfy
\begin{eqnarray}
\label{Eq2}
v\overline{v} = \overline{v} v =1, & n \overline{n} = \overline{n} n, & vn=qnv,
\nonumber \\
n \overline{v} = q \overline{v} n, & v \overline{n}=q \overline{n} v, &
\overline{n} \, \overline{v} = q \overline{v} \, \overline{n},\nonumber \\
\Delta (n)=n \otimes \overline{v} + v \otimes \overline{n}, & \Delta (v) = v
\otimes v, \nonumber \\
\Delta (\overline{n}) = \overline{n} \otimes v + \overline{v} \otimes \overline
{n}, & \Delta (\overline{v}) = \overline{v} \otimes \overline{v},\\
\epsilon (n) = \epsilon (\overline{n}) = 0, & \epsilon (v) = \epsilon (
\overline{v}) =1, \nonumber \\
S(n)=-q^{-1}n,& S(v)=\overline{v},\nonumber \\
S(\overline{n}) = -q \overline{n}, & S(\overline{v})=v, \nonumber
\end{eqnarray}
with $\overline{q}=q$.  (These relations can also be obtained through
contraction of $SU_{q}(2)$, as described in the next section.)

For the calculations which follow, it is convenient to introduce the operators
$\theta$, $\overline{\theta}$, $m$, and $\overline{m}$, defined by
\begin{eqnarray}
& v=e^{\frac{i}{2} \theta},\:\: \overline{\theta}=\theta ,\: \: m=nv, \:\:
\overline{m} = \overline{v} \,\overline{n}. &
\end{eqnarray}
In this basis, the coproducts take on the particularly nice form
\begin{eqnarray}
\Delta (m) =& m \otimes 1 + e^{i \theta} \otimes m, & \Delta ( \overline{m}) =
\overline{m} \otimes 1 + e^{-i \theta} \otimes \overline{m},\nonumber \\
\Delta (\theta)  =& \theta \otimes 1 + 1 \otimes \theta.
\end{eqnarray}

\noindent {\bf Remark:}  The matrix $E$ given by
\begin{equation}
E = \left( \begin{array}{ll}
e^{i \theta} & m \\ 0 & 1
\end{array} \right)
\end{equation}
satisfies the relations
\begin{eqnarray}
\Delta (E) = E \dot{\otimes} E, & S(E) = E^{-1}, & \epsilon (E) = I.
\end{eqnarray}
These are exactly the relations one would expect for an element of a quantum
group.  Notice that the action of $E$ on the column vector $\left(
\begin{array}{c}
z \\ 1
\end{array} \right)$, where $z$ is a complex coordinate, is given by
\begin{eqnarray}
z \mapsto e^{i \theta} z + m, & \overline{z} \mapsto e^{-i \theta} \overline{z}
+ \overline{m}.
\end{eqnarray}
We may therefore identify $E$ as an element of the deformed 2-dimensional
Euclidean group $E_q(2)$.

$Fun(E_{q}(2))$ is the algebra of all $C^{\infty}$ functions in the group
parameters of $E_{q}(2)$, i.e. the algebra spanned by ordered monomials in
$\theta$, $m$, and $\overline{m}$.  Thus, $Fun(E_{q}(2))$ is taken to be
$span \{ \theta ^{a} m^{b} \overline{m}^{c} \mid a,b,c =0,1,... \}$.  The dual
to $Fun(E_{q}(2))$ is $U_{q}e(2)$, the quantized universal enveloping algebra
of
the 2-dimensional Euclidean algebra.  We take $\xi$, $\mu$, and $\nu$ to be the
elements of $U_{q}e(2)$ which give
\begin{eqnarray}
<\mu, \theta ^{a} m^{b} \overline{m} ^{c}>=\delta_{a0} \delta_{b1} \delta_{c0},
& <\nu,\theta ^{a} m^{b} \overline{m} ^{c}>= \delta_{a0} \delta_{b0}
\delta_{c1}, \nonumber \\
<\xi,\theta ^{a} m^{b} \overline{m} ^{c}>=\delta_{a1} \delta _{b0} \delta_{c0}.
\end{eqnarray}
Using the coproduct on $Fun(E_{q}(2))$ to obtain the multiplication on $U_{q}e(
2)$ gives
\begin{eqnarray}
<\nu ^{k} \mu ^{l} \xi ^{n},\theta ^{a} m^{b} \overline{m}^{c}>=[k]_{q}!
[l]_{q^{-1}}! n! \delta_{na} \delta_{lb} \delta_{kc}, & [x]_{q}! \equiv
{\displaystyle \prod_{y=1}^{x} \frac{q^{2y}-1}{q^{2}-1}},
\end{eqnarray}
so $\{ \nu^{k} \mu^{l} \xi ^{n} \mid k,l,n=0,1,... \}$ is a basis for
$U_{q}e(2)$.  The rest of the Hopf algebra structure of $U_{q}e(2)$ can be
similarly obtained:
\begin{eqnarray}
[\xi,\mu]=i\mu,& [\xi,\nu]=-i\nu,& \mu \nu = q^{2}\nu \mu,
\nonumber \\
\Delta (\mu)= \mu \otimes q^{2i \xi} + 1 \otimes \mu, & \Delta (\nu) = \nu
\otimes q^{2i \xi} + 1 \otimes \nu, \nonumber \\
\Delta (\xi) = \xi \otimes 1 + 1 \otimes \xi, & \epsilon (\mu)=
\epsilon (\nu) = \epsilon (\xi) =0, \\
S(\mu)= -\mu q^{-2i \xi}, & S(\nu) = -\nu q^{-2i \xi}, & S(\xi) = -\xi.
\nonumber
\end{eqnarray}

If $H$ is a *-Hopf algebra whose coproduct and antipode satisfy
\begin{eqnarray}
\Delta (\overline{h}) = \overline{\Delta (h)}, & \overline{S(\overline{h})
}=S^{-1}(h)
\end{eqnarray}
for all $h \in H$, then an involution on $H^{*}$ can be defined by \cite{DSWZ}
\begin{equation}
\label{dual}
<\overline{\chi},h>=<\chi,S^{-1}(\overline{h})>^{*},
\end{equation}
where $\chi \in H^{*}$.  Using this to define complex conjugation
on $U_{q}e(2)$ gives
\begin{eqnarray}
\overline{\xi} = - \xi, & \overline{\mu} = -q^{2}\nu, & \overline{\nu} =
-q^{-2}\mu.
\end{eqnarray}
Defining new operators $J$, $P_{+}$, and $P_{-}$ as
\begin{eqnarray}
J \equiv i\xi, & P_{+} \equiv q q^{-i \xi}\nu, & P_{-} \equiv -q^{-1} \mu q^{-i
\xi},
\end{eqnarray}
gives $\overline{J}=J$, $\overline{P_{\pm}}=P_{\mp}$, and
\begin{eqnarray}
\label{eq2}
[J,P_{\pm}]=\pm P_{\pm}, & [P_{+},P_{-}]=0, \nonumber \\
\Delta (P_{\pm})=P_{\pm} \otimes q^{J} + q^{-J} \otimes P_{\pm}, & \Delta (J) =
J \otimes 1 + 1 \otimes J, \\
\epsilon (P_{\pm}) = \epsilon (J) =0, \nonumber \\
S(J) = -J, & S(P_{\pm})=-q^{\pm 1} P_{\pm}. \nonumber
\end{eqnarray}

\section{$U_{q}e(2)$ from $SU_{q}(2)$}
\label{Peter}

In this section we will show how the deformed Euclidean group $E_{q}(2)$ can
be obtained from $SU_{q}(2)$ by contraction and how this implies a similar
contraction scheme for the deformed Lie algebra $U_{q}su(2)$, giving
an independent derivation of $U_{q}e(2)$.

\subsection{$E_{q}(2)$ by contraction of $SU_{q}(2)$}
\label{Peter1}

Recall \cite{RTF}, \cite{Z} the commutation relations for $SU_{q}(2)$, which
may be written in compact matrix notation as
\begin{eqnarray}
\label{RTT}
R_{12}T_{1}T_{2} = T_{2}T_{1}R_{12}, & det_{q}T = 1, & T^{\dagger} = T^{-1},
\nonumber \\
\Delta (T) = T \dot{\otimes} T, & \epsilon (T) = I, & S(T)=T^{-1},
\end{eqnarray}
where
\begin{eqnarray}
T =  \left( \begin{array}{lr}
\alpha & -q \overline{\gamma} \\ \gamma & \overline{\alpha}
\end{array} \right), & R = q^{-1/2} \left( \begin{array}{cccc}
q & 0 & 0 & 0 \\
0 & 1 & 0 & 0 \\
0 & \lambda & 1 & 0 \\
0 & 0 & 0 & q
\end{array} \right),
\end{eqnarray}
and $\lambda = q-q^{-1}$.  Now set $\alpha \equiv v$, $\overline{\alpha}
\equiv \overline{v}$, $\gamma \equiv l \overline{n}$ and $\overline{\gamma}
\equiv ln$, where $l \in {\bf R} \backslash \{0\}$ is the contraction
parameter.
 Written in terms of $v$, $\overline{v}$, $n$ and $\overline{n}$, relations
(\ref{RTT}) become
\begin{eqnarray}
&det_{q}T=v \overline{v} + q^{2} l^{2} n \overline{n} =
\overline{v}v +  l^{2} \overline{n} n =1,& \nonumber \\
&n \overline{n} = \overline{n} n,\,\, v n = q n v, \,\,
v \overline{n} = q \overline{n} v,\,\, \mbox{etc.} & \nonumber
\end{eqnarray}
and coincide with (\ref{Eq2}) in the limit $l \rightarrow 0$, i.e.
$E_{q}(2)$ is a contraction of $SU_{q}(2)$.

\subsection{$U_{q}e(2)$ by contraction of $U_{q}su(2)$}
\label{Peter2}

The deformed universal enveloping algebra $U_{q}su(2)$, dual to $Fun(SU_{q}(2))
$, is generated by hermitean operators $H$, $X_{+}$, $X_{-}$ satisfying
\begin{eqnarray}
\label{HXX}
[H,X_{\pm}] = \pm 2 X_{\pm}, & [X_{+},X_{-}] = \frac{q^{H}-q^{-H}}{q-q^{-1}},
\nonumber \\
\Delta (H) = H \otimes 1 + 1 \otimes H, & \Delta (X_{\pm}) =
X_{\pm} \otimes q^{H/2} + q^{-H/2} \otimes X_{\pm}, \nonumber \\
\epsilon (H)= \epsilon(X_{\pm})=0, \\
S(H) = -H, & S(X_{\pm}) = -q^{\pm 1} X_{\pm}. \nonumber
\end{eqnarray}
Following \cite{RTF} these relations can be rewritten as
\begin{eqnarray}
\label{RLL}
R_{12}L^{\pm}_{2}L^{\pm}_{1}=L^{\pm}_{1}L^{\pm}_{2}R_{12},&
R_{12}L^{+}_{2}L^{-}_{1}=L^{-}_{1}L^{+}_{2}R_{12},\nonumber \\
\Delta (L^{\pm}) = L^{\pm} \dot{\otimes} L^{\pm}, & \epsilon(L^{\pm})=I, \\
S(L^{\pm})=(L^{\pm})^{-1}, \nonumber
\end{eqnarray}
where $L^{\pm}$ are given by
\begin{eqnarray}
L^{+} =  \left( \begin{array}{lr}
q^{-H/2} & q^{-1/2} \lambda X_{+} \\ 0 & q^{H/2}
\end{array} \right),&  L^{-} = \left( \begin{array}{lr}
q^{H/2} & 0 \\ -q^{1/2} \lambda X_{-} & q^{-H/2}
\end{array} \right).
\end{eqnarray}
Using this matrix notation, there is an elegant way of stating the duality
between the group and the algebra by means of the commutation relations
\begin{eqnarray}
\label{TRL}
L^{+}_{1} T_{2} = T_{2} R_{21} L^{+}_{1}, & L^{-}_{1} T_{2} = T_{2} R^{-1}_{12}
L^{-}_{1},
\end{eqnarray}
as described in \cite{Z}.  Equations (\ref{TRL}) are consistent with the inner
products
\begin{eqnarray}
<L^{+}_1,T_{2}>=R_{21}, & <L^{-}_{1},T_{2}>=R^{-1}_{12},
\end{eqnarray}
given in \cite{RTF}.  Furthermore, equations (\ref{RLL}) can be derived as
consistency conditions to (\ref{RTT}) and (\ref{TRL}).  In addition, complex
conjugation can be defined as an involution
on the extended algebra generated by products of
$T$ and $L^{\pm}$. This agrees with (\ref{dual}).  Unitarity of $T$ then
implies
$(L^{+})^{\dagger}= (L^{-})^{-1}$, i.e. $\overline{H} = H$, $\overline{X_{\pm}}
=X_{\mp}$.

In the present case equations (\ref{TRL}) become
\begin{eqnarray}
\label{ga}
Hv=vH-v, & X_{+}v=q^{1/2}vX_{+}-lqnq^{H/2}, & X_{-}v=q^{1/2}vX_{-},
\nonumber \\
lH \overline{n}=l(\overline{n}H-\overline{n}), &
lX_{+} \overline{n} = q^{1/2} \overline{n}lX_{+} + \overline{v}q^{H/2}, &
lX_{-} \overline{n}= lq^{1/2}\overline{n}X_{-},
\end{eqnarray}
plus the complex conjugate relations.

The way that the deformation parameter $l$ appears in these relations suggests
the definition of new operators $P_{+} \equiv lX_{+}$, $P_{-} \equiv
\overline{P_{+}} = lX_{-}$ and $J \equiv H/2$, so that we will retain
non-trivial commutation relations for $P_{\pm}$ and $J$ with
$v$, $\overline{v}$, $n$ and $\overline{n}$ in the limit $l \rightarrow 0$.
Inserting $P_{\pm}$ and $J$ into equation (\ref{HXX})  we again
obtain $U_{q}e(2)$ (see (\ref{eq2}))
as a contraction of $U_{q}su(2)$ in this limit.

\section{Conclusion}

Through equation (\ref{ga}) the contraction on the group level
(section~\ref{Peter1}) motivates a particular contraction scheme at the
algebra level (section~\ref{Peter2}).
The two methods outlined in sections~\ref{sec-Paul} and~\ref{Peter} are
summarized in the following (commutative) diagram:

\begin{picture}(300,70)
\put(90,50){$SU_{q}(2)$}
\put(210,50){$E_{q}(2)$}
\put(90,0){$U_{q}su(2)$}
\put(210,0){$U_{q}e(2)$}
\put(105,45){\vector(0,-1){30}}
\put(225,45){\vector(0,-1){30}}
\put(130,54){\vector(1,0){80}}
\put(132,4){\vector(1,0){76}}
\put(155,56){\tiny contraction}
\put(162,48){\tiny $l \rightarrow 0$}
\put(155,6){\tiny contraction}
\put(162,-2){\tiny $l \rightarrow 0$}
\put(106,30){\tiny dual}
\put(226,30){\tiny dual}
\end{picture}
\vspace{2 mm}

Note that the {\em algebra} obtained in (\ref{eq2}) is the same as the
classical
2-dimensional Euclidean algebra $e(2)$ (with $P_{\pm}=P_{x}\pm i P_{y}$ and
$J$ as hermitean generators) \cite{CGST}.  Note, however, as a {\em Hopf
algebra} it is still deformed; the deformation parameter $q$ remains unchanged.

\end{document}